\documentclass[pre,aps,amsmath,amssymb,reprint]{revtex4-1}
\usepackage{graphicx}
\usepackage{dcolumn}
\usepackage{bm}
\usepackage{color}
\usepackage{multirow}

\renewcommand{\langle}{\left<}
\renewcommand{\rangle}{\right>}

\renewcommand{\hat}[1]{{\widehat #1}}
\newcommand{\comments}[1]{}

\begin{document}


\title{Bistable Intrinsic Charge Fluctuations of a Dust Grain Subject to Secondary Electron Emission  in a Plasma }



\author{B. Shotorban}\email{ babak.shotorban@uah.edu}
\affiliation{Department of Mechanical and Aerospace Engineering, The University of Alabama in Huntsville, Huntsville, Alabama 35899}



\begin{abstract}

A master equation was formulated to study intrinsic charge fluctuations of a grain in a plasma as ions and primary electrons  are attached to the grain through collisional collection, and secondary electrons are emitted from the grain.  Two different plasmas with Maxwellian and non-Maxwellian distributions  were considered.  The  fluctuations could be bistable in either plasma when the secondary electron emission is present, as two stable macrostates, associated with two stable roots of the charge net current, may exist.  Metastablity of fluctuations,  manifested by the passage of the grain charge between two macrostates, was shown to be possible. 

\end{abstract}


\maketitle 


\section{Introduction}
\label{Introduction}

Various mechanisms, including ion and electron collisional collection and resulting secondary emission of ions or electrons,  contribute to the charging of a dust grain in a plasma.  Since the collision of plasma particles with the grain occurs at random times, the net electric charge possessed by the grain fluctuates in time even if the plasma parameters such as temperature and number densities are  fixed.   This kind of  fluctuations, which take place in systems with discrete particles,  are known as {\em intrinsic noise} \cite{Kampen07}. The intrinsic noise cannot be switched off as it is inherent in the actual physical mechanism, e.g., electron or ion electron collision or emission in grain charging mechanism, which is responsible for the evolution of the system. Intrinsic charge fluctuations  refer to random variation of the grain charge by this intrinsic noise.   

Description of  intrinsic charge fluctuations of grains was the subject of a number of  studies \cite{CG94,MR95,MRS96,MR97,gordiets1998charge,KNPV99,gordiets1999charge,shotorban2011nonstationary,asgari2011stochastic,shotorban2012stochastic,matthews2013cosmic,shotorban2014intrinsic,asgari2014non,mishra2015statistical}.  
\citet{CG94} studied the fluctuations through a Monte Carlo approach and concluded that they are more important for smaller grains. The grain charge is correlated with the grain size so this conclusion is consistent with that the net elementary charge $Z$ possessed by the grain should have fluctuations with  $Z_\mathrm{rms}\propto \sqrt{|\langle Z \rangle|}$, which suggested by \citet{morfill1980dust}.  \citet{CG94}  also showed that the fluctuating charge of small grains could experience positive values. It is known that the  grain mean charge at equilibrium is negative because in an average sense, the grain collects mobile electrons more than  ions,  as it approaches to an equilibrium charge. \citet{MR95} proposed a one-step process master equation~\cite{Kampen07} for the grain charge density function, then derived a Fokker-Planck equation for it and  showed that if the condition $e^2/4\pi \epsilon_0 Rk_BT_e \ll1$, where $R$ is the radius of the grain and $T_e$ is the electron temperature, is satisfied, the charge distribution at stationary states is Gaussian with an average and variance  correlated to the ion and electron currents to  the grain.  
Defining the system size as $\Omega=4\pi\epsilon_0Rk_BT_e/e^2$,  \citet{shotorban2011nonstationary} derived a Gaussian solution at non-stationary states for the Fokker-Planck equation formulated through the system size expansion of the master equation \cite{Kampen07}.  
In the non-stationary state Gaussian solution, the rate of the mean grain charge   correlates with the rate of charge to the net current. This mean equation  is the macroscopic equation \cite{Kampen07} of the grain charging system, and it is the same equation widely used for the grain charging with negligible fluctuations, which is the conservation of the grain charge.   The rate of the grain charge variance is correlated with the currents and their derivatives evaluated  at  the charge mean. \citet{shotorban2014intrinsic} lately extended this model to include multi-component plasmas where there are various kinds of singly- or multiply-charged negative or positive ions and showed that the grain charge distribution still follows Gaussianity  when $\Omega$ is sufficiently large.  In all the references  discussed above, collisional collections of electrons and ions were the only mechanism of charging.  \citet{gordiets1998charge} obtained an analytical solution for the PDF at stationary states for a master equation that included the effect of the electron detachment, e.g., secondary electron emission (SEE), assuming that the grain charge does not experience positive values, i.e, a half-infinite range assumption $Z=0, -1, -2, \ldots$.  This kind stationary-state solution is unique for  the master equation of a general one-step process with a half-infinite or finite range of the variable whereas it is  not unique for a range consisting of all integers \cite{Kampen07}.  Later, \citet{gordiets1999charge} formulated an improved version of the master equation that they had originally proposed \cite{gordiets1998charge}, relaxed the half-infinite range assumption, and derived an approximate analytical solution for the PDF at the stationary state.  This approximation is not well justified for grains where the PDF varies substantially over a small range of charges.  \citet{KNPV99} studied the effects of thermionic emission and UV irradiation, separately, while electron collisional collection was present. They concluded that $Z_\mathrm{rms}\propto \sqrt{|\langle Z \rangle|}$ is valid for these situations as well. 
Lately, \citet{mishra2015statistical}  studied the fluctuations in multi-component plasma through a population balance equation resembling the master equation. They also included the influence of photoemission from dust through irradiation by laser light in their study.  It is noted that all  above but \citet{asgari2011stochastic,asgari2014non}  used  Markov approaches to describe grain charge fluctuations.  

The most well known effect of SEE on grain charging is a  bifurcation phenomenon: two identical grains in an identical plasma environment may have two different stable  charge values, one positive and one negative \cite{meyer1982flip,horanyi1998electrostatic}. Thus, a small variation in the parameters may cause a rapid change from one equilibrium charge to another.  Interestingly, \citet{lai1991theory} showed that in the spacecraft charging, the bifurcation phenomena caused by SEE could involve three stable equilibrium charge values.  The experimental study of \citet{walch1995charging} on charging of grains with energetic electrons that resulted in secondary electron emission, showed the distribution of grain charge could be bimodal.  However, they asserted  that the lack of a unique value may be due to fluctuations in the plasma parameters or small differences in the grains.  

The current study is on the influence of SEE on  grain charge intrinsic fluctuations with a focus on bistability. Bistability  occurs in stochatic systems with two stable macrostates \cite{Kampen07,Gardiner04}. The bistability of the grain charging system is associated with the bifurication phenomena described above.  The fluctuations in a bistable system may be metastable \cite{Kampen07}, where the fluctuations are at one macrostate for a while and at a random time,  a passage to the other macrostate takes place and at a random time, the system returns  to the first macrostate and this cycle continues. Whether  grain charge fluctuations could be metastable is investigated in this work. 
In section \ref{sec:MathemticalFormulation}, first, a master equation describing  the fluctuations of the grain charge in the presence of SEE mechanism is presented, and then currents of ions, primary electrons and secondary electrons of a Maxwellian plasma and a non-Maxwellian plasma are shown.   In section \ref{sec:results}, results are shown and discussed. Conclusions are made in section \ref{sec:conclusions}.

\section{Mathematical Formulation}
\label{sec:MathemticalFormulation}

Assuming that the charging of the grain undergoes a Markov process, the following master equation can be formulated for the probability density function of the grain charge $P(Z,t)$:

\noindent

\begin{eqnarray}
{dP(Z,t)\over dt}&=&({\mathbb E}-1)f_0(Z)I_e(Z)P(Z)\nonumber\\
&+&\sum_{n=1}^{M-1} \left({\mathbb E}^{-n}-1\right)f_{n+1}(Z)I_e(Z)P(Z)\nonumber\\
&+&\left({\mathbb E}^{-1}-1\right)I_i(Z)P(Z),
\label{eq:oneStepMaster}
\end{eqnarray}


\noindent where ${\mathbb E}$ is an operator defined by ${\mathbb E}^kg(Z)=g(Z+k)$ for any integer number $k$, $n$ indicates the number of secondary electrons emitted from the grain upon the impact of one primary electron, $M$ is the maximum number of secondary electrons that can be emitted,  $I_i(Z)$ and $I_e(Z)$ are the currents of ions and primary electrons to the grain, respectively, and $f_n(Z)$ is the probability distribution of emission of $n$ electrons in a single incident of a primary impact, i.e., the fraction of primary electrons that result in the emission of $n$ secondary electrons in one single attachment incident. Hence, the rate of the attachment of the primary electrons that do not cause secondary emission is $f_0(Z)I_e(Z)$, 
and $f_n(Z)I_e(Z)$ indicate the rate of the attachment of the primary electrons that cause the emission of  $n$ secondary electrons in one incident. The jump process associated with  eq.~(\ref{eq:oneStepMaster}) is regarded to that the attachment of a primary electron to the grain causes the emission of $n$ secondary electrons; thus the net change of the grain charge is $n-1$. In other words, $Z(t)$, the charge of the grain at time $t$, jumps to $Z(t)+n-1$. It is noted that the master equation of \citet{gordiets1999charge} is a special case of eq.~(\ref{eq:oneStepMaster}) with $M=3$. Also, two following special cases of the master equation~(\ref{eq:oneStepMaster}) regarded as one-step processes are worth noting:

\begin{itemize}
\item $M=0$, which corresponds to a case where no SEE occurs, i.e.,  $f_0(Z)=1$ and  $f_n(Z)=0$ for $n>0$. In this case, the second term on the right hand side of eq.~(\ref{eq:oneStepMaster}) vanishes and the master equation of the grain charing is retrieved \cite{MR95,shotorban2011nonstationary},

\item $M=1$, which corresponds a case that at most one secondary electron is emitted  so  $f_0(Z)<1$ and $f_1(Z) =1-f_0(Z)$. 
\end{itemize}

Defining the system size $\Omega$ as a reference constant charge number and having changed the variable $Z=\Omega\phi(t)+\Omega^{1/2}\xi$, where $Z$ is  modeled by a combination of a deterministic part $\phi(t)$ scaled by $\Omega$, and a random part $\xi$ scaled by $\Omega^{1/2}$, a macroscopic  equation associated with eq.~(\ref{eq:oneStepMaster}) can be derived through the system size expansion method~\cite{Kampen07,shotorban2014intrinsic}:

\begin{equation}
\frac{d\phi}{dt}=a_1(\phi),
\label{eq:macrostate}
\end{equation} 

\noindent
where $a_1(\phi)=\Omega^{-1}I_n(\Omega \phi)$, $I_n(.)= I_i(.)-I_e(.)+I_s(.)$ is the net current to the grain, and $I_s(.)$ is the SEE current to the grain. A solution of the macroscopic eq.~(\ref{eq:macrostate}) is a time-dependent macrostate of the grain charging system while  the solution of $a_1(\phi)=0$ is a stationary macrostate of the system~\cite{Kampen07}. 

\citet{Kampen07} classifies the stable, bistable, and unstable stochastic systems through $a_1(\phi)$ in eq.~(\ref{eq:macrostate}): A stochastic system is stable when $a_1'(\phi)<0$ where $ a_1'(\phi)\equiv da_1/d\phi$; it is bistable when there are two stable stationary macrostates, i.e., there are two solutions for $a_1(\phi)=0$ and at the vicinities of them, $a_1'(\phi)<0$ holds; and it is unstable when $a_1'(\phi)>0$. The bistability classification here is in harmony with the bifurcation phenomenon  based on the roots of the net current, which is reviewed in sec.~\ref{Introduction}. In other words, having neglected the fluctuations of $Z$, i.e., $\xi=0$, one obtains $Z=\Omega \phi$ and eq. (\ref{eq:macrostate}) is readily simplified to $dZ/dt=I_n(Z)$, which can be used to find the time evolution of $Z$. If an initial $Z$ is within the domain of attraction of a stable root of $I_n(Z)=0$, then $Z$ approaches it at the stationary state when there are no fluctuations.   Each root is associated with one stationary macrostate of the system and the system stability defined above can be likewise determined through the sign of $I_n'(Z)$. When the fluctuations of the grain charge are taken into account,  there is a probability for a fluctuation  to carry the charge from the domain of attraction of one root to another.  This situation is investigated in next section.

The SEE current is correlated with $I_e(Z)$ and $f_n(Z)$, and this correlation is found through  the mean secondary electron yield defined by

\begin{equation}
{\overline n}(Z) =I_s(Z)/I_e(Z).
\label{eq:IsIe}
\end{equation}  

\noindent
 In addition, ${\overline n}(Z)$ is correlated with $f_n(Z)$ through the definition of the mean ${\overline n}(Z) =\sum_{n=1}^M n f_n(Z)$. Using these two equations and the normalization condition, i.e., $\sum_{n=0}^M f_n(Z) =1$, one obtains
\begin{equation}
I_s(Z)=\left[ 1- f_0(Z)+\sum_{n=2}^M (n-1) f_n(Z) \right ]I_e(Z).
\end{equation}

\begin{figure}[t]
\begin{center}
\includegraphics[width=.8\columnwidth, angle=0]{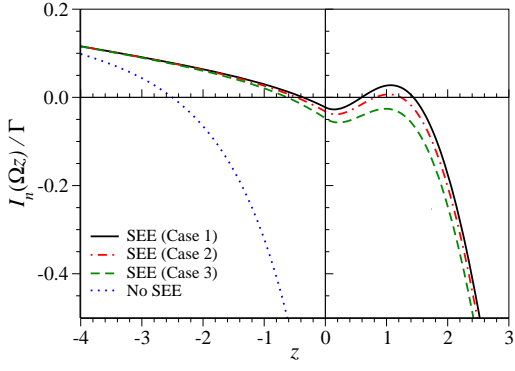}
\end{center}
\caption{Dimensionless net current versus dimensionless grain charge $z=Z/\Omega$, where $\Omega=4\pi\epsilon_0Rk_BT_e/e^2$ and $\Gamma=\pi R^2n_e  \sqrt{8k_BT_e/\pi m_e }$, for $T_s/T_e=1.5$ in a Maxwellian plasma~\cite{meyer1982flip}; Case 1:~$\delta_M=15$ and $E_M/4kT_e=45.6$; Case 2:~$\delta_M=14.85$ and $E_M/4kT_e=45.6$;    Case 3:~$\delta_M=15$ and $E_M/4kT_e=47$. }
\label{fig:netCurrent}
\end{figure}

Here,  a binomial distribution is proposed for $f_n(Z)$:
\begin{equation}
f_n(Z)=\binom{M}{n}p^n(1-p)^{M-n},
\label{eq:binomial}
\end{equation}
where for $M>0$,  $p={\overline n}(Z)/M$ where  ${\overline n}(Z)$ is given in eq.~(\ref{eq:IsIe}).  Binomial distributions are used for the Monte Carlo modeling of SEE in the electron-surface collision \cite{furman2002probabilistic}. For $M=1$ in eq.~(\ref{eq:binomial}), $f_0(Z)=1-{\overline n}(Z)$ and hence, $f_0(Z)I_e(Z)P(Z)=[I_e(Z)- I_s(Z)]P(Z)$.  For $M=1$, the summation terms in this equation are zero.  It is noted that $I_e(Z)- I_s(Z)$ is the  electron net current to the grain so the charging process here is modeled as each primary electron impact incident causing no or one secondary electron emission at most.  In eq. (\ref{eq:binomial}), a sufficient condition for the positivity of $f_n(Z)$ is $1-p>0$, which is equivalent to 
\begin{equation}
{\overline n}(Z)<M. 
\label{eq:condition}
\end{equation}
This inequality sets the requirement for the minimum  $M$ according to variation of ${\overline n}(Z)$ versus $Z$.

Following \citet{meyer1982flip}, who investigated the bifurcation phenomena associated with SEE, both Maxwellian and non-Maxwellian plasmas are considered here:

For  Maxwellian plasmas, it can be shown   \cite{Kimura1998,shotorban2014intrinsic}

\begin{equation}
I_e(Z)=  \Gamma\times \left\{ 
\begin{array}{l l}
  1+{Z\over \Omega} & \quad Z\ge 0,\\\\
\exp\left ({Z\over \Omega} \right)& \quad Z< 0,\\ \end{array} \right.
\label{eq:elecCurrent}
\end{equation}

\begin{equation}
I_i(Z)=  \Gamma\hat{n}_i \sqrt{\hat{T}_i\over \hat{m}_i}\times \left\{ 
\begin{array}{l l}
  1-{Z\over \hat{T}_i\Omega} & \quad Z\le 0,\\\\
\exp\left (-{Z\over \hat{T}_i\Omega} \right)& \quad Z> 0,\\ \end{array} \right.
\label{eq:ionCurrent}
\end{equation}

\noindent
where $n_i$ and $T_i$ are  the number density and temperature of ions, respectively. Also, $\hat{T}_i=T_i/T_e$, $\hat{m}_i=m_i/m_e$, $\hat{n}_i=n_i/n_e$,

\begin{equation}
\Omega={{4 \pi \epsilon_0Rk_BT_e}\over e^2 },
\label{eq:omega}
\end{equation}
\begin{equation}
\Gamma=\pi R^2n_e  \sqrt{8k_BT_e\over\pi m_e } ={\Omega \omega_\mathrm{pe}R
\over\sqrt{2\pi}\lambda_{De}},
\label{eq:Gamma}
\end{equation}

\noindent
where $\lambda_{De}=\sqrt{\epsilon_0k_BT_e/n_ee^2}$ is the electron Debye length and 
$\omega_{pe}=\sqrt{n_e e^2/\epsilon_0m_e}$ is the electron plasma frequency. 

The SEE current is obtained by \cite{meyer1982flip} 

\begin{widetext}
\begin{equation}
I_s(Z)=  3.7\delta_M\Gamma\times \left\{ 
\begin{array}{l l}
  \left(1+{Z\over{\Omega\hat{T}_s}}\right)\exp\left(-{Z\over{\Omega\hat{T}_s}}+{Z\over\Omega}\right )F_{5,B}\left( {E_M\over{4k_BT_e}}\right)& \quad Z\ge 0,\\\\
\exp\left ({Z\over \Omega} \right)F_5\left( {E_M\over{4k_BT_e}}\right)& \quad Z< 0,\\ \end{array} \right.
\label{eq:SEE}
\end{equation}
\end{widetext}

\onecolumngrid

\begin{figure}{b}
\begin{tabular}{cc}
\includegraphics[width=.35\columnwidth, angle=0]{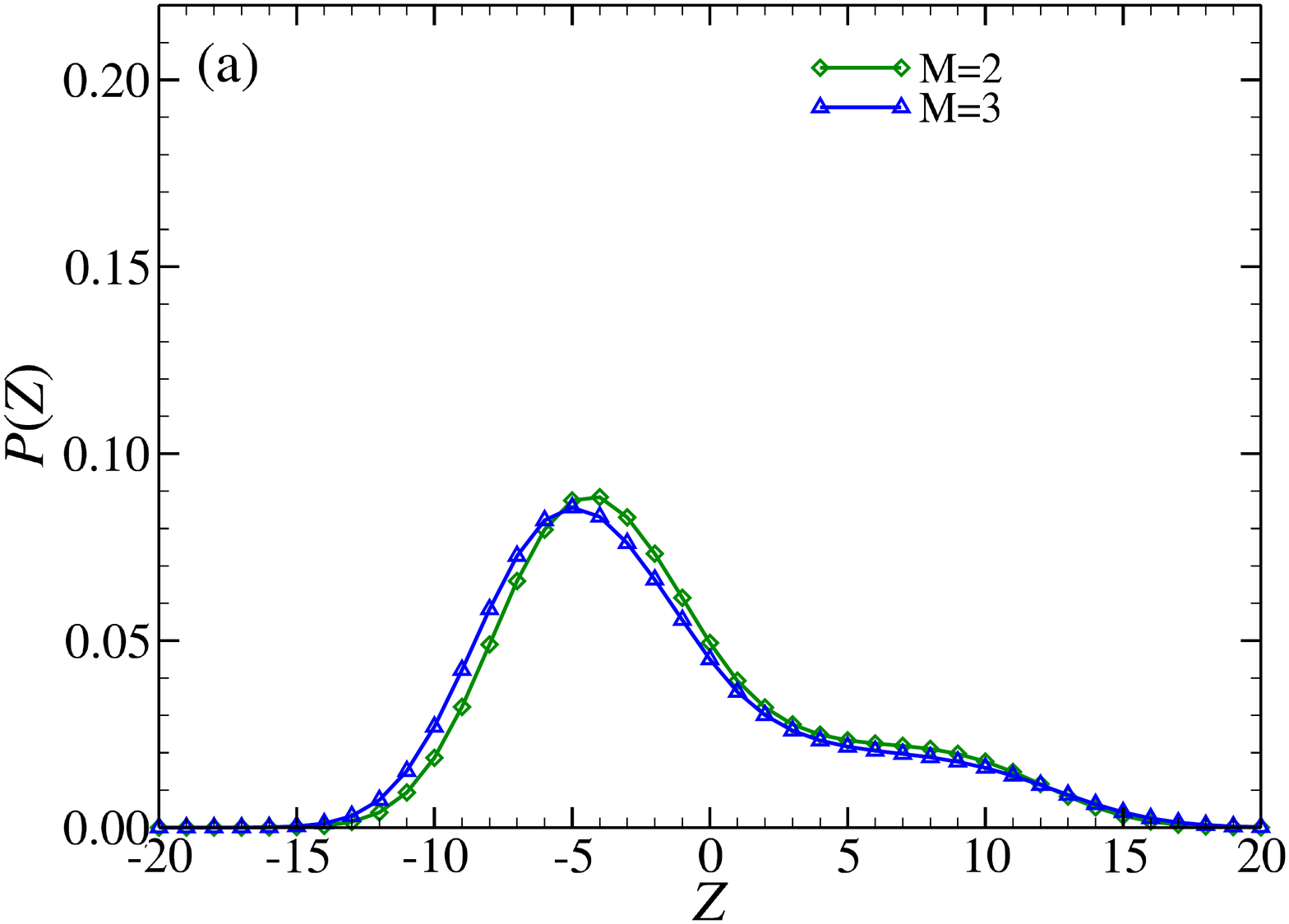} &
\includegraphics[width=.35\columnwidth, angle=0]{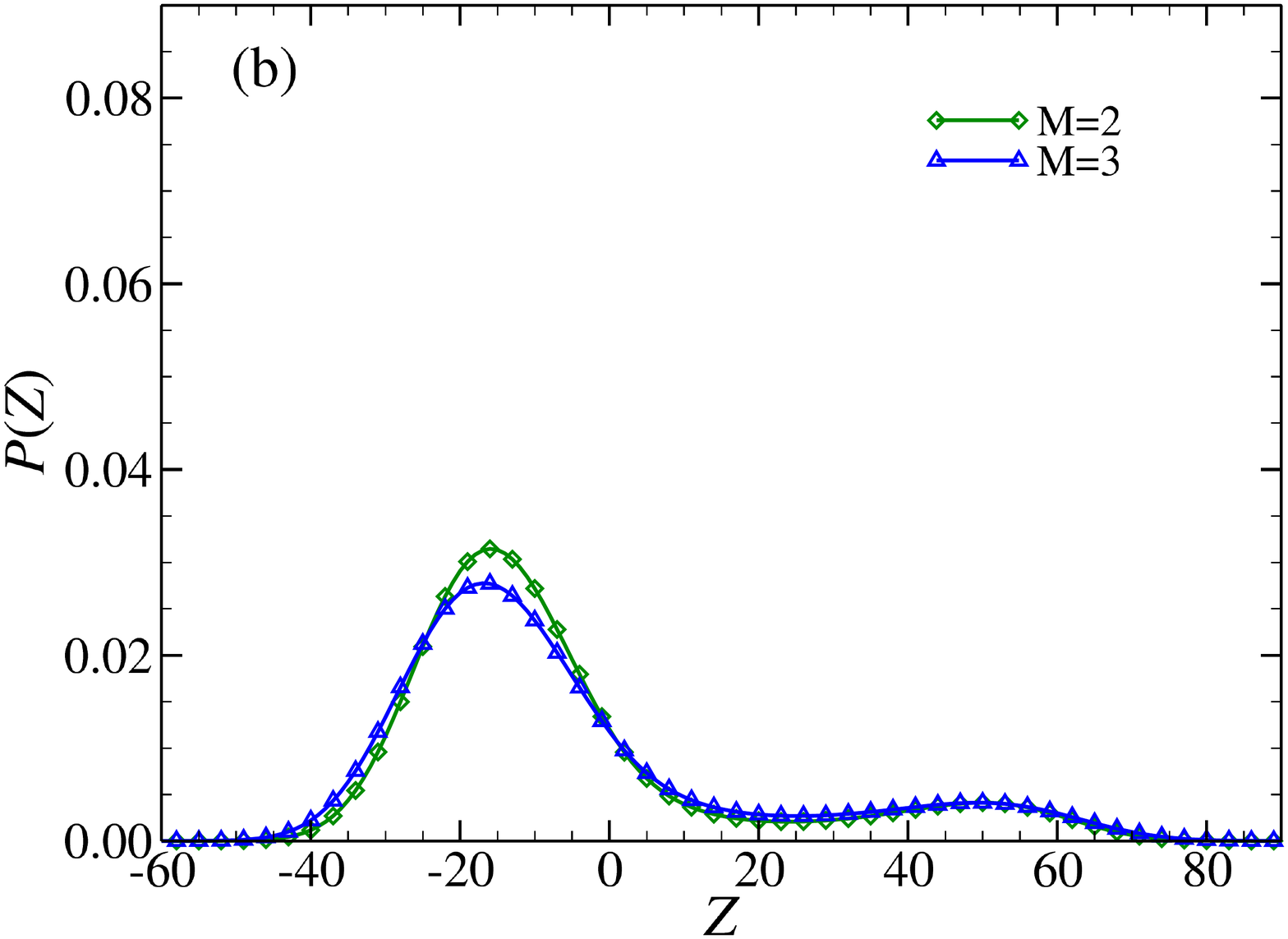} \\

\includegraphics[width=.35\columnwidth, angle=0]{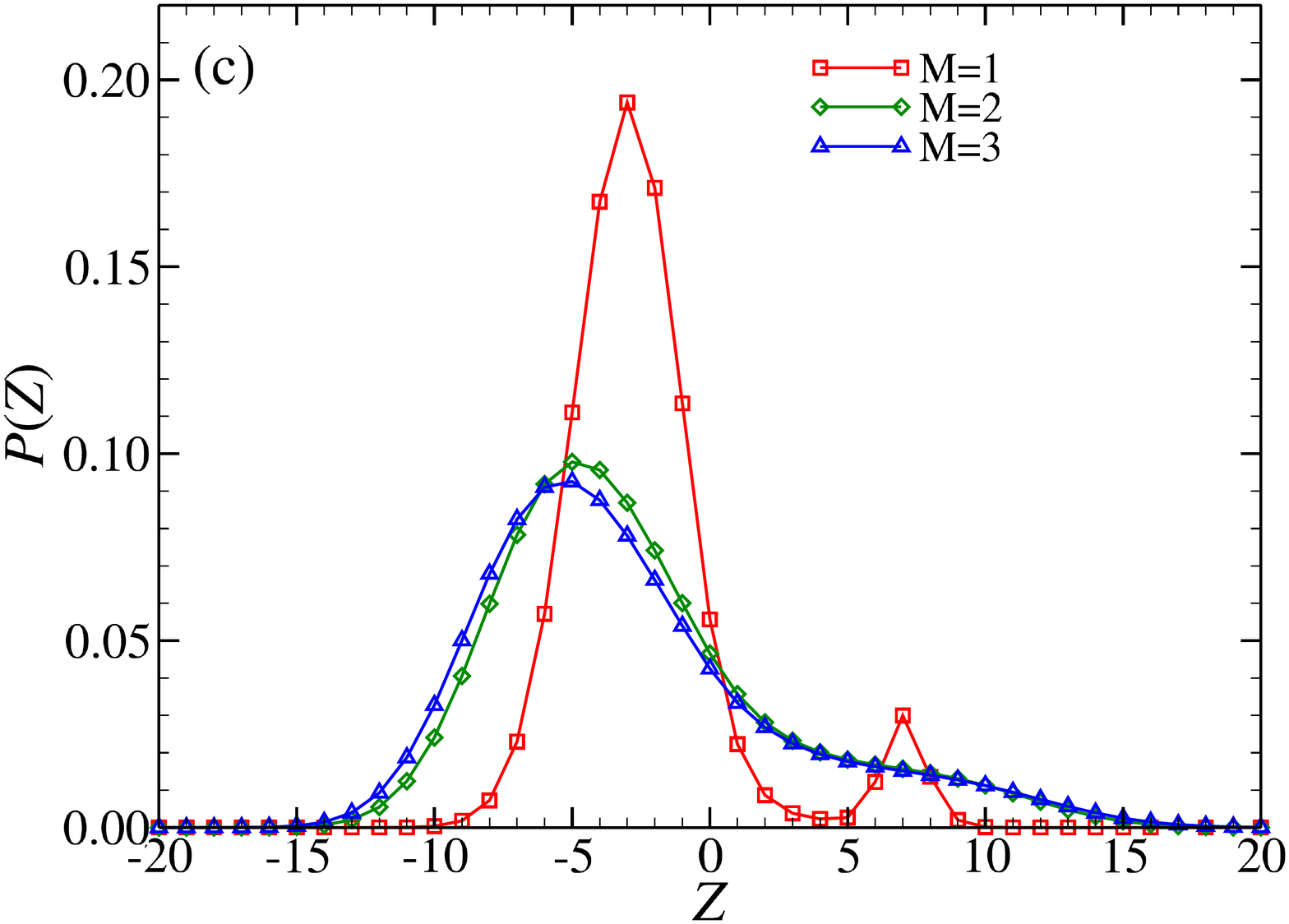} &
\includegraphics[width=.35\columnwidth, angle=0]{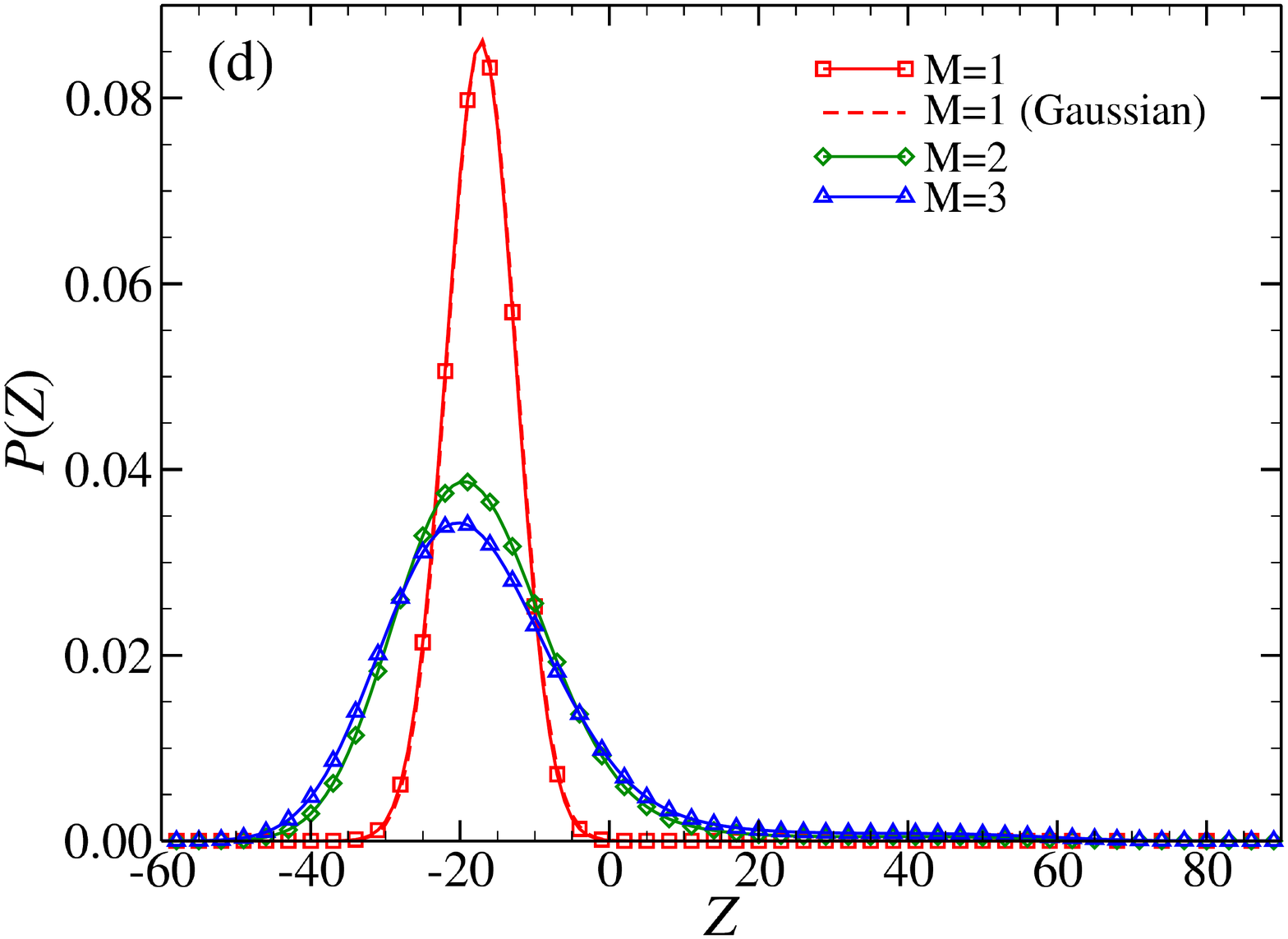}\\

\includegraphics[width=.35\columnwidth, angle=0]{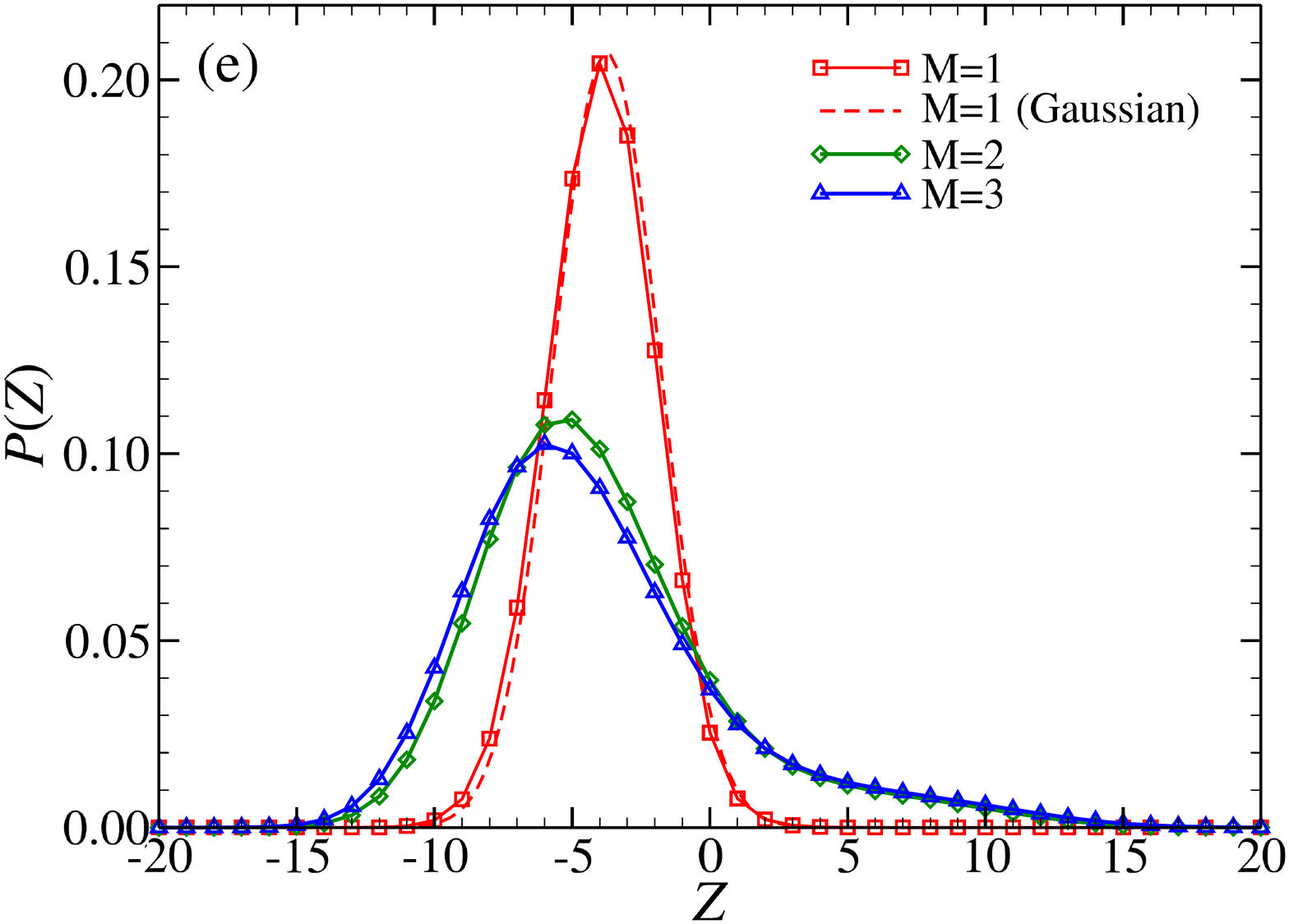} &
\includegraphics[width=.35\columnwidth, angle=0]{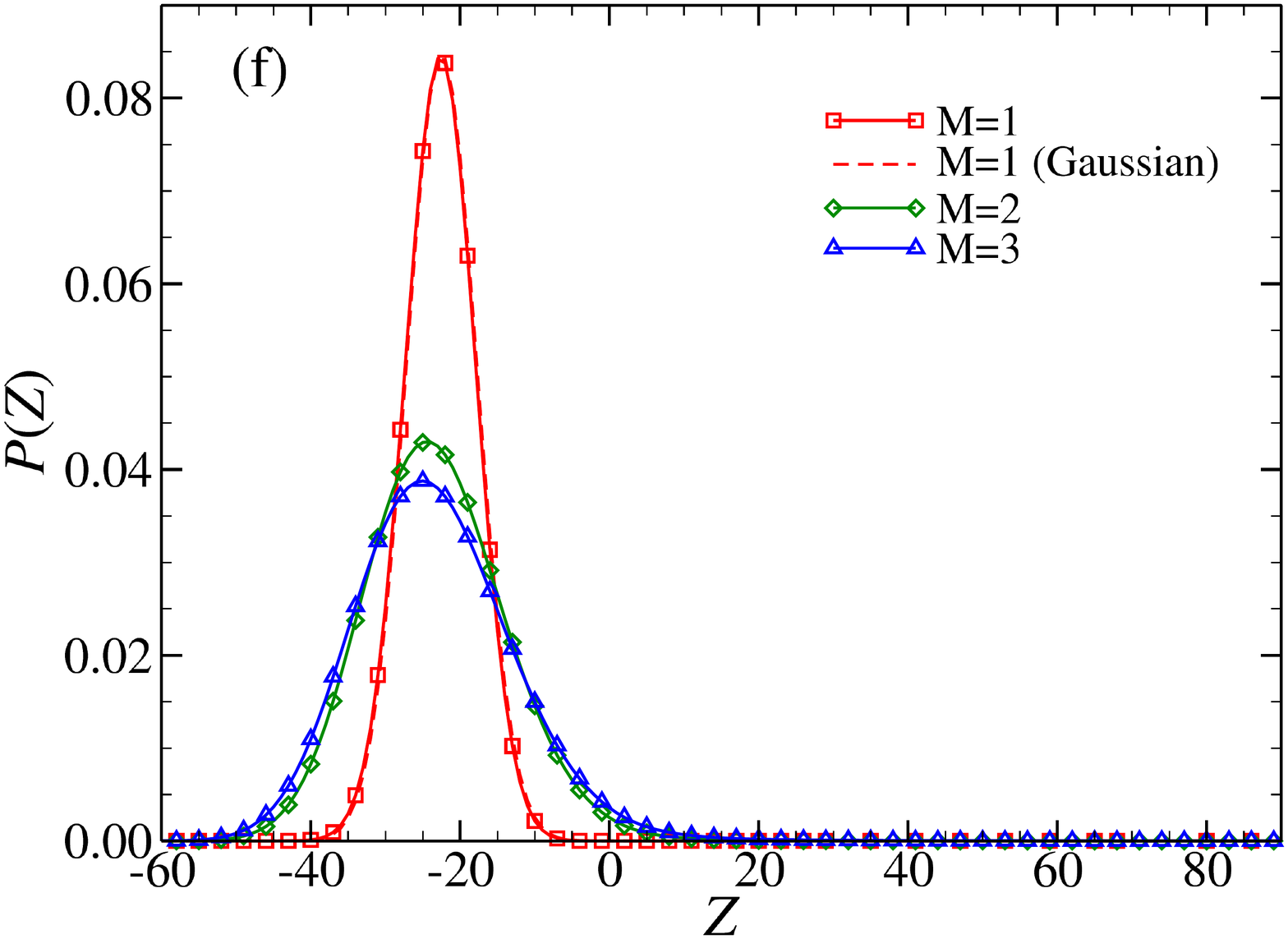}\\

\end{tabular}
\caption{Probability density function of  grain charge in a Maxwellian plasma; (a) Case 1 with $R=5$nm; (b) Case 1 with $R=30$nm; (c) Case 2 with $R=5$nm; (d) Case 2 with $R=30$nm; (e) Case 3 with $R=5$nm; (f) Case 3 with $R=30$nm. Gaussian solutions are obtained by the system size expansion method \cite{shotorban2014intrinsic,shotorban2011nonstationary}. See the caption of fig.~\ref{fig:netCurrent} for parameters associated with Case 1-3.}
\label{fig:pdfMaxwellian}
\end{figure}

\twocolumngrid

\noindent 
where 
\[ F_5(x)=x^2\int_0^\infty  u^5\exp\left(-xu^2-u\right)du,\] 
\[F_{5,B}(x)=x^2\int_B^\infty  u^5\exp\left(-xu^2-u\right)du,\] 
where $B=\sqrt{{4k_BT_e}Z/\Omega E_M}$.

Electrons in the non-Maxwellian plasma are assumed to have a bi-Maxwellian distribution. The electron current is obtained by adding an identical term, where $n_e$ and $T_e$ are to be replaced by $n_H$ and $T_H$,  to the right side of eq. (\ref{eq:elecCurrent}) and changing $n_e$ to $n_e-n_H$ in the first term \cite{meyer1982flip}.

\section{Results and Discussions}
\label{sec:results}

Dimensionless net current is plotted against dimensionless grain potential in fig.~\ref{fig:netCurrent} for a Maxwellian plasma~\cite{meyer1982flip}. Seen in this figure is that when the SEE mechanism is lacking, the net current curve crosses the horizontal axis only at one point so there is only one root.  The system is stable in this case as $I_n'(Z)<0$ for all values of $Z$. This negativity is due to that the attaching electrons are more mobile than attaching ions.  On the other hand, it is seen in the figure that when the SEE mechanism is present, the net current may have up to three roots, one negative and two positive (Cases 1 and 2).  The root at the middle is unstable whereas two others are stable so the grain charge fluctuations are bistable in Cases 1 and 2.  In case 3,  only one stable root exists and for all values of $Z$ except the domain restricted between local maxima and minima, the system is stable.  It is noted that the SEE cases seen in fig.~\ref{fig:netCurrent} are different through small changes made in the SEE current parameters $\delta_M$ or $E_M$. A detailed study on the impact of these parameters and plasma parameters on the roots of net current can be found in Ref.~\onlinecite{meyer1982flip}. 

\begin{figure}
\begin{tabular}{c}
\vspace{-1.5in}

\includegraphics[width=\columnwidth, angle=0]{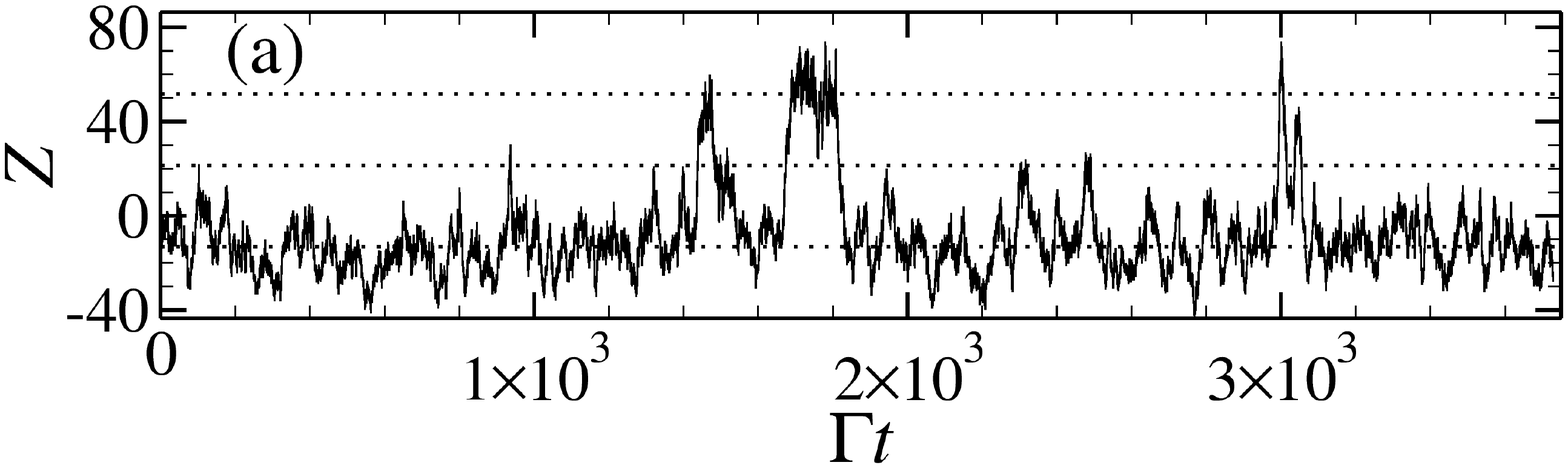} \\

\vspace{-1.5in}

\includegraphics[width=\columnwidth, angle=0]{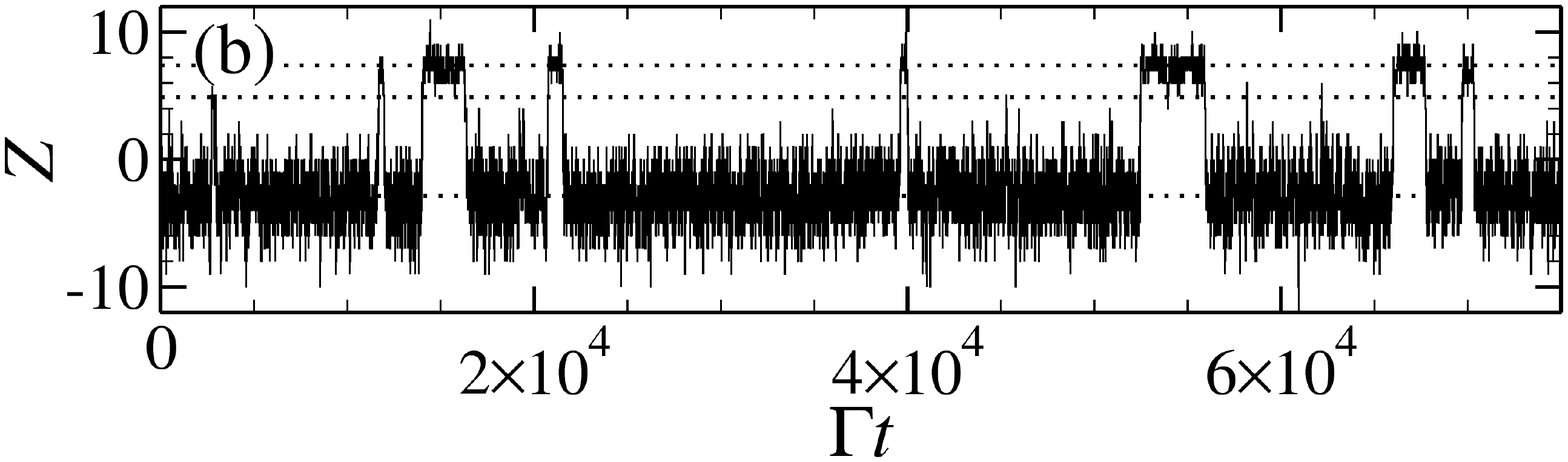} \\

\vspace{-1.5in}

\includegraphics[width=\columnwidth, angle=0]{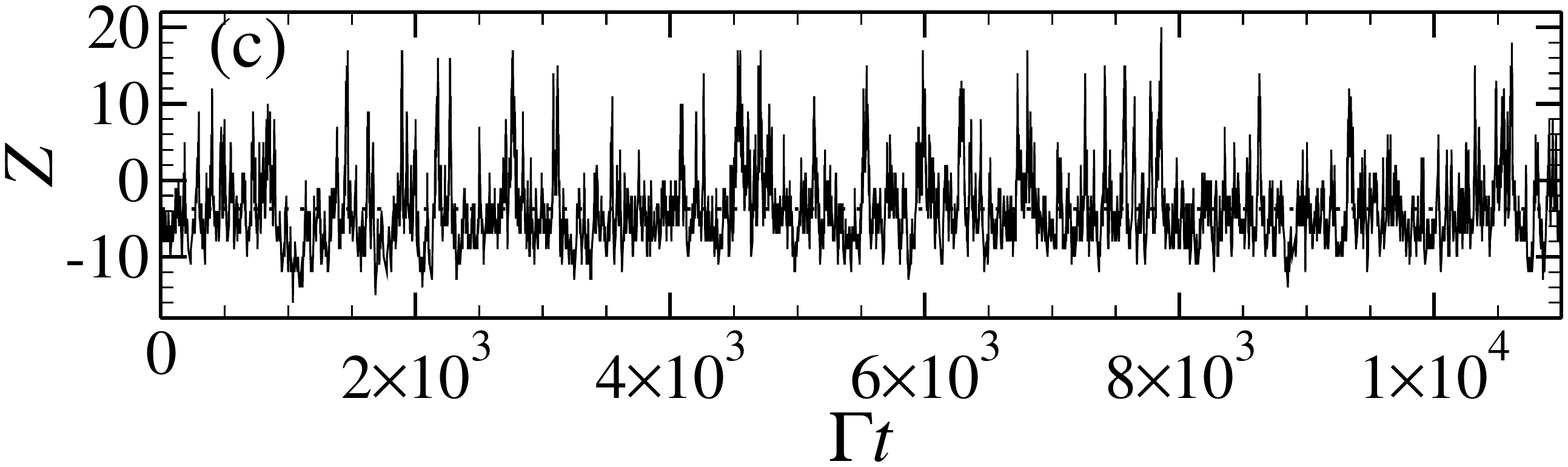} \\

\end{tabular}

\caption{Grain charge variation in time in a Maxwellian plasma with SEE; (a) Case 1 with $R=30$nm and $M=3$; (b) Case 2 with $R=5$nm and $M=1$; (c) Case 3 with $R=5$nm and $M=3$. The dotted lines show the roots of the net current. }
\label{fig:chargeTimeVariationMaxwellian}
\end{figure}

The PDF of the grain charge obtained through a numerical solution of the master equation (\ref{eq:oneStepMaster}) for a Maxwellian plasma (Case 1-3 illustrated in fig.~\ref{fig:netCurrent}) for two grain sizes $R=5$ and $R=30$nm are shown in fig.~\ref{fig:pdfMaxwellian}. A prominent deviation from Gaussian distribution is observed for most of the cases. However, the PDF was verfied to be very close to Gaussian  for $M=1$ in figs. \ref{fig:pdfMaxwellian}(d-f) as compared to a Gaussian solution obtained by the system size expansion method \cite{shotorban2014intrinsic,shotorban2011nonstationary}. For $R=5$nm and $M=1$, seen in fig.~\ref{fig:pdfMaxwellian}(c), the PDF is bimodal, i.e., with two distinct local maxima, and for two other values of $M$, it is not.  It is  borne in mind  that for all cases in figs \ref{fig:pdfMaxwellian}(a-d), there are two stable roots of the net current so they all are considered bistable according to the classification in the previous section.  The bimodal PDF is also seen in  fig.~\ref{fig:pdfMaxwellian}(b) for a larger grain with $R=30$nm.  However, when the same SEE parameters are used for smaller grain $R=5$nm, no bimodal distribution is observed (see fig.~\ref{fig:pdfMaxwellian}a). No bimodal distribution is observed in fig.~\ref{fig:pdfMaxwellian}(e-f) which is for the SEE cases with only one root of the net current.  Although, in these two subfigures, the deviation of the distribution from Gaussianity is substantial for $M=2$ and 3.

\begin{figure}
\begin{center}
\includegraphics[width=.8\columnwidth, angle=0]{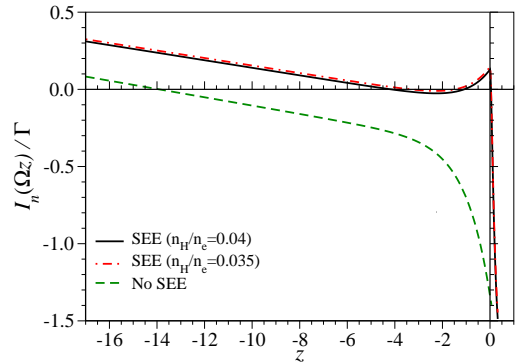}
\end{center}
\caption{Dimensionless net current versus dimensionless grain charge $z=Z/\Omega$ in a bi-Maxwellian plasma; $T_e=T_i=25$eV, $\delta_M=3$, $E_M/k_BT_e=16$, $T_s/T_e=1$,  $T_H/T_e=100$, and $M=3$ \cite{meyer1982flip}. }
\label{fig:nonMaxNetCurrent}
\end{figure}

Figure \ref{fig:chargeTimeVariationMaxwellian} displays   time histories of  grain charges.   The discrete stochastic method \cite{shotorban2014intrinsic}, adapted from Gillespie's algorithm \cite{Gillespie76,gillespie2007stochastic}), is utilized to simulate  the grain charge fluctuations governed by the master equation (\ref{eq:oneStepMaster}).  Time histories seen in figs. \ref{fig:chargeTimeVariationMaxwellian}(a,b),   are for the  bistable cases shown in fig. \ref{fig:pdfMaxwellian}(b) for $M=3$ and fig. \ref{fig:pdfMaxwellian}(c) for $M=1$, respectively. The fluctuations in these two cases are characterized by two distinct time scales: one associated with fluctuations around either of stable roots of the net current, i.e., charging macrostates, and the other associated with the spontaneous switches between them. A system with this behavior is called metastable~\cite{Kampen07}. A  transition from the macrostate associated with the negative stable charge to the other macrostate  is attributed to a sequence of incidents most of which increase the grain charge by one or two elementary charges. These incidents could be the attachment of an ion or the attachment of a primary electron that results in the emission of two or more  of secondary electrons. On the other hand, a transition from the macrostate associated with the positive stable root of the net current to the other macrostate  is attributed to a sequence of incidents most of which are the attachments of a primary electron without  emitting  a secondary electron.  Fig.~\ref{fig:chargeTimeVariationMaxwellian}(c) which corresponds to the PDF shown in fig. \ref{fig:pdfMaxwellian}(e) with $M=3$ is not considered bistable as  the net current in this case has only one root. 

Shown in fig.~\ref{fig:nonMaxNetCurrent} is the net current variation against the grain charge in a bi-Maxwellian 
plasma. For the shown SEE cases, there are two negative and one positive roots for the net current so the system is bistable in both SEE cases. The positive root is very close to the origin of the coordinates and the net current  has a very sharp variation around this root.  Although the curves of the SEE cases shown in this figure seem very similar, they are different as the negative roots in the case with  $n_H/n_e=0.035$ are slightly closer to each other than the  case with  $n_H/n_e=0.04$. The root of the net current in the No SEE case is somewhat far from the roots of the SEE cases.

\begin{figure}
\begin{center}
\includegraphics[width=.8\columnwidth, angle=0]{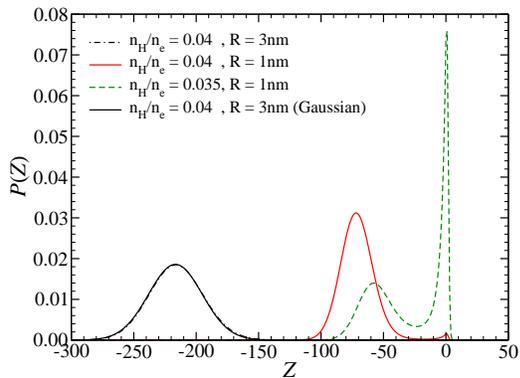}
\end{center}
\caption{Probability density function of grain charge in a bi-Maxwellian plasma. See the caption of fig. \ref{fig:nonMaxNetCurrent} for parameters. }
\label{fig:nonMaxPDF}
\end{figure}

\begin{figure}[t]
\begin{tabular}{c}
\vspace{-1.5in}
\includegraphics[width=\columnwidth, angle=0]{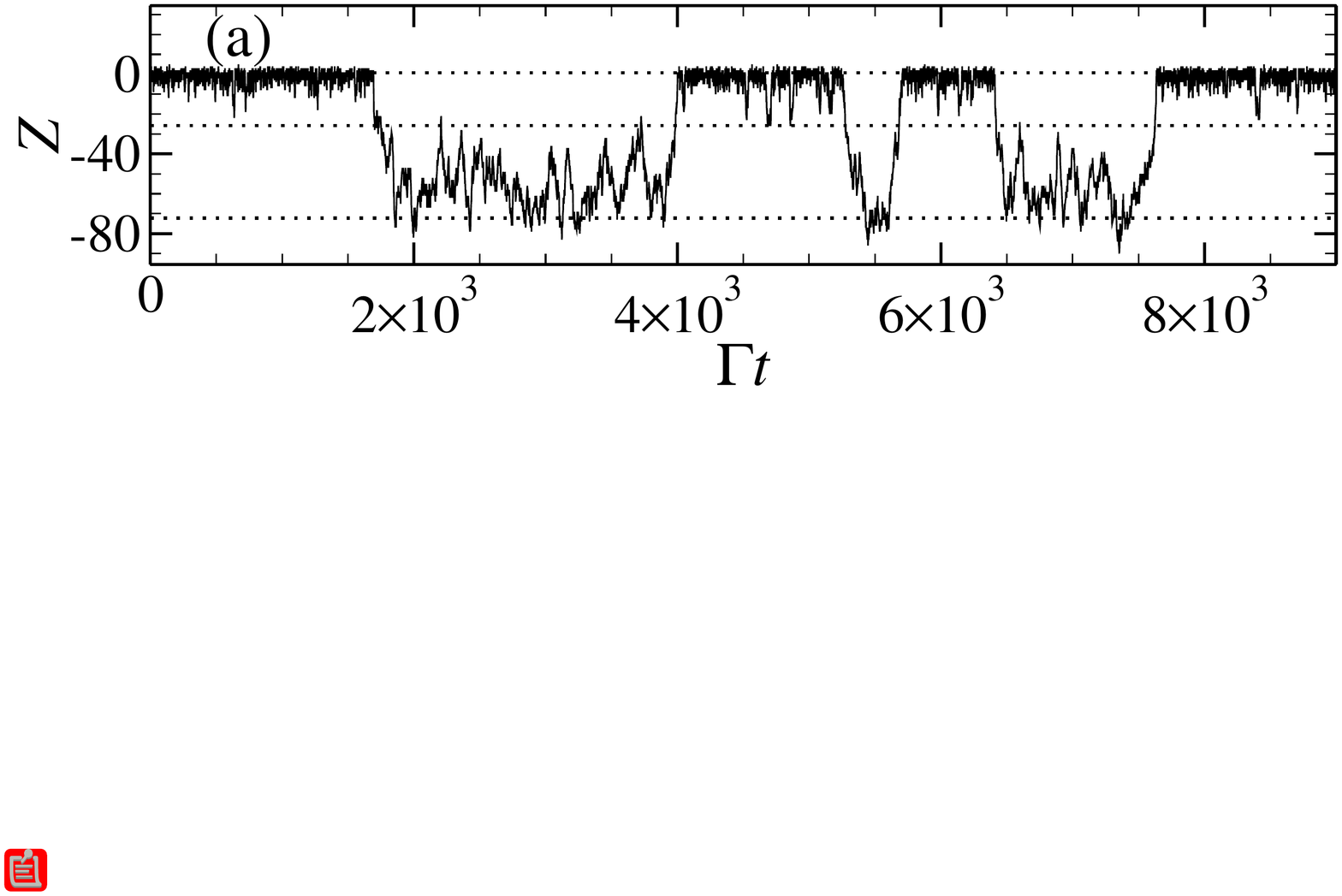} \\
\vspace{-1.5in}
\includegraphics[width=\columnwidth, angle=0]{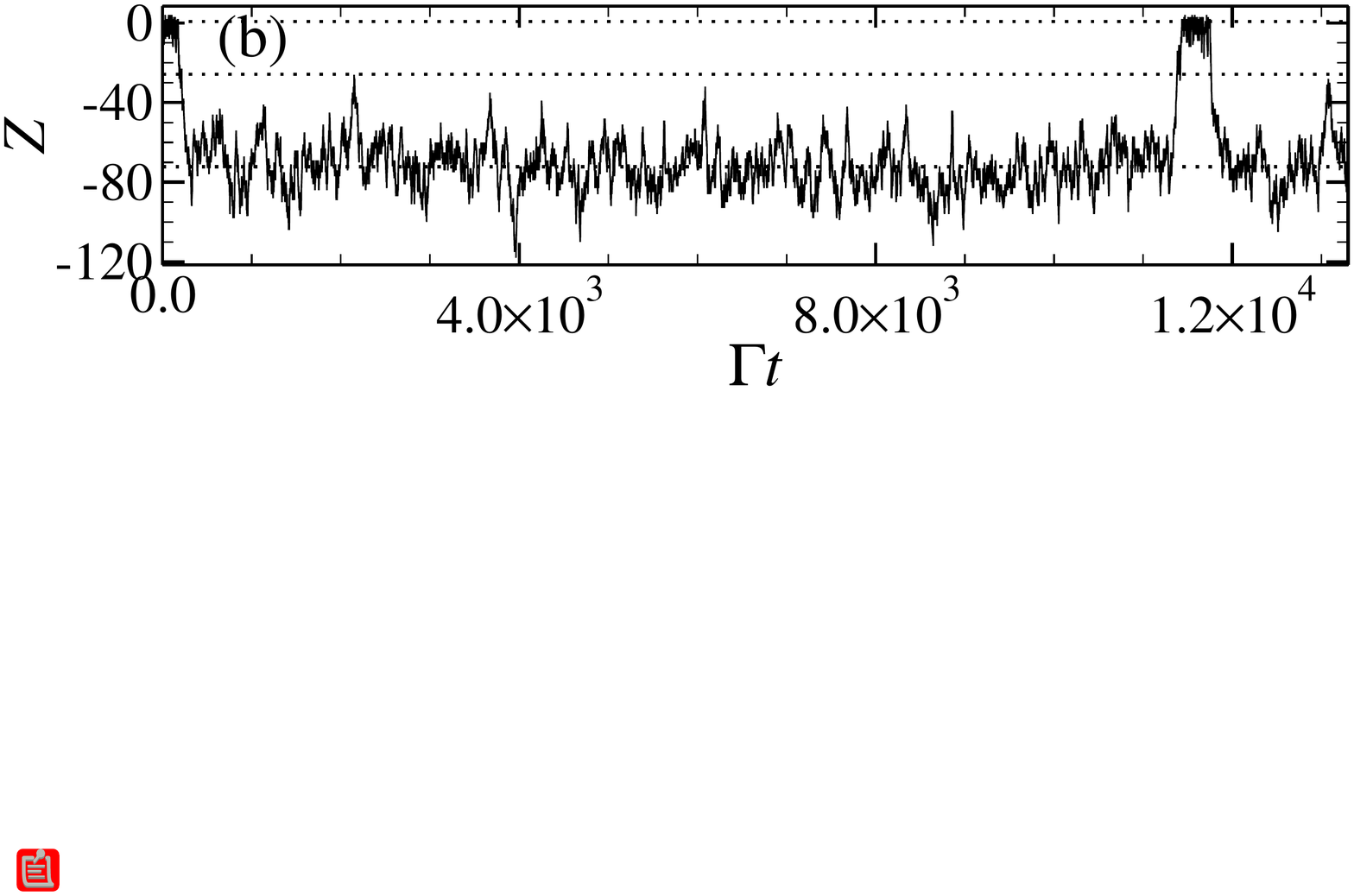} \\

\end{tabular}

\caption{Grain charge fluctuations  in a bi-Maxwellian plasma with SEE, $M=3$, and  initial charge $Z(0)=0$; (a) $n_H/n_e=0.035$ and $R=1$nm; (b) $n_H/n_e=0.04$ and $R=1$nm; (c) $n_H/n_e=0.04$ and $R=3$nm. }
\label{fig:chargeTimeVariationNonMaxwellian}
\end{figure}

Figure \ref{fig:nonMaxPDF} displays the grain charge PDF in the studied bi-Mawellian plasma for two grain sizes. All three cases shown in this figure are associated with the SEE cases in fig.~\ref{fig:nonMaxNetCurrent}, which are  bistable. A bimodal distribution is observed for smaller grain with $R=1$nm at both $n_H/n_e=0.035$ and $0.04$.  Although, the difference between these two values of $n_H/n_e$  is around \%13, the bimodal forms of their associated PDF's are very different.  The peak value of the PDF seen at around $Z=0$ for $n_H/n_e=0.035$ is at least an order of magnitude larger than  that for $n_H/n_e=0.04$. For this case, the value of the left peak is  an order of magnitude larger that the right mode.  For the grain with a larger radius $R=3$nm, no bimodal behavior is observed. For this case, also, a Gaussian solution is obtained by the system size expansion with an initial condition $\langle Z(0)\rangle/\Omega=-3$. An excellent agreement between the Gaussion solution and the master equation solution is observed.  When $\langle Z(0)\rangle=0$ is used, the solution at the stationary state is a sharp Gaussian function at around $Z=0$. Time history of the grain charge is shown in fig.~\ref{fig:chargeTimeVariationNonMaxwellian} with panels (a) and (b) associated with solid- and dashed-line PDFs, respectively,  in fig.~\ref{fig:nonMaxPDF}. An obvious metastability is observed for these two cases.

\section{Summary and Conclusions}
\label{sec:conclusions}
A master equation was formulated to include the effect of secondary electron emission in addition to collisional attachment of ions and electrons on the intrinsic charge fluctuations of a grain. Grain charging in both Maxwellian and non-Maxwellian plasmas were considered. In both plasmas, the fluctuations could be bistable, as the system could have two stable macrostates. In the absence of SEE mechanism, the bistabillity is not possible as the system always have a single macrostate.  It was shown that if the system is bistable, the grain charge can be metastable. That is a situation where  the fluctuations are characterized by two distinct time scales - one associated with the fluctuations at either macrostate and the other associated with the spontaneous transition between mactrostates. A  switch from the macrostate associated with the negative stable root of the net current to the macrostate associate with the positive stable root of the net current  is attributed to a sequence of incidents almost all of which increase the grain charge by one or two elementary charge. On the other hand, a converse switch  is attributed to a sequence of incidents most of which are the attachments of primary electrons without  resulting in the emission of  secondary electrons. 

%
%

\acknowledgements

The author acknowledges the support by the National Science Foundation through award PHY-1414552.



%



\end{document}